\documentclass{LMCS}

\def\dOi{11(1:6)2015}
\lmcsheading%
{\dOi}
{1--18}
{}
{}
{Nov.~11, 2013}
{Mar.~10, 2015}
{}

\ACMCCS{[{\bf Theory of computation}]: Logic---Proof theory\,/\,Logic
  and verification}

\usepackage{graphicx}
\usepackage{amsmath,amssymb, stmaryrd}

\usepackage{bussproofs}
\usepackage{comment}

\usepackage{hyperref}

\usepackage{booktabs}

\newcommand{\Nat}{{\mathbb N}}

\newcommand{\compatible}[2]{\mathrm{compatible}(#1,#2)}
\newcommand{\incompatible}[2]{\mathrm{incompatible}(#1,#2)}

\newcommand{\var}{\mathrm{Var}}
\newcommand{\NC}{\mathrm{nc}}
\newcommand{\forallnc}{\forall_{\NC}}
\newcommand{\existsnc}{\exists_{\NC}}
\newcommand{\ire}[2]{#1\,\mathbf{r}\,#2}
\newcommand{\mybar}[1]{\overline{#1}}

\newcommand{\grec}{\mathrm{gRec}}

\newcommand{\Boolelim}[3]{\mathrm{if}\ #1\ \mathrm{then}\ #2\ \mathrm{else}\ #3}
\newcommand{\inhab}[1]{\mathrm{inhab}_{#1}}

\newcommand{\True}{\mathrm{True}}
\newcommand{\muu}[3]{\mu(#1;#2;#3)}
\newcommand{\measure}[2]{|#2 \setminus\!\!\setminus \var (#1)|}
\newcommand{\weight}[1]{\#(#1)}

\newcommand{\dpllrule}[1]{\mathbf{#1}}
\newcommand{\Conflict}{\dpllrule{Conflict}}
\newcommand{\Unit}{\dpllrule{Unit}}
\newcommand{\Elim}{\dpllrule{Elim}}
\newcommand{\Red}{\dpllrule{Red}}
\newcommand{\Split}{\dpllrule{Split}}
\newcommand{\resrule}[1]{\mathbf{#1}}
\newcommand{\Res}{\resrule{Res}}
\newcommand{\Sub}{\resrule{Sub}}

\newcommand{\modres}[1]{\overset{#1}{\underset{\text{\tiny Res}}{\vdash}}}
\newcommand{\moddpll}[1]{\overset{#1}{\underset{\text{\tiny DPLL}}{\vdash}}}

\newcommand{\consvals}{\mathrm{Cons}}

\begin{document}
\title[Extracting verified decision procedures: 
DPLL and Resolution]{Extracting verified decision procedures: \\
DPLL and Resolution}

\author[U.~Berger]{Ulrich Berger\rsuper a}	
\address{{\lsuper{a,b,d}}Swansea University, UK}	
\email{\{u.berger,csal,m.seisenberger\}@swansea.ac.uk}  

\author[A.~Lawrence]{Andrew Lawrence\rsuper b}	

\author[F.~Nordvall Forsberg]{Fredrik Nordvall Forsberg\rsuper c}	
\address{{\lsuper c}University of Strathclyde, UK}	
\email{fredrik.nordvall-forsberg@strath.ac.uk}  

\author[Seisenberger]{Monika Seisenberger\rsuper d}	



\begin{abstract} 
This article is concerned with the application of the program extraction technique to a new class of problems: the synthesis of decision procedures for the classical satisfiability problem that are correct by construction.
To this end, we formalize a completeness proof for the DPLL proof system and
extract a SAT solver from it.  When applied to a propositional formula in conjunctive normal form the program produces either a satisfying assignment or a DPLL derivation showing its unsatisfiability. We use non-computational quantifiers to remove redundant computational content from the extracted program and translate it into Haskell to improve performance. We also prove the equivalence between the resolution proof system and the DPLL proof system with a bound on the size of the resulting resolution proof.
This demonstrates that it is possible to capture quantitative
information about the extracted program on the proof level.
The formalization is carried out in the interactive proof assistant Minlog.
\end{abstract}

\maketitle

\section{Introduction}

In order for verification tools to be used in an industrial context
they have to be trusted to a high degree and in many cases are
required to be certified.  We present a new application of program
extraction to develop a formally verified decision procedure for the
satisfiability problem for propositional formulae in conjunctive normal
form.  The procedure is based on the DPLL proof system \cite{DP,DPLL}
which is also the basis of most contemporary SAT solvers that are used
in an industrial context.

The need for verified SAT solvers is obvious; they are part of
safety critical software, and also used for the verification and 
certification thereof.  SAT solvers are nowadays highly
optimized for speed, which makes the introduction of errors 
(in the process of optimization) more
likely, and their verification more difficult.  Besides the
correctness 
also totality (or universality) of
SAT solvers is an issue. For example, in the 2012 SAT competition
(\texttt{www.smtcomp.org}) many systems were not total in the sense
that they returned ``Unknown'' for certain inputs signifying that they
could not deal with the given problem.

In this paper we report about the extraction of a SAT solver that is both
correct and total by construction. In addition, it produces in the
unsatisfiable case a formal proof of this fact, which is recognized in
the SAT community as a highly desirable feature of SAT solvers.
To be more precise, we formalize a correctness and completeness proof of
the DPLL proof system in the interactive theorem prover
Minlog, and use Minlog's program
extraction facilities to obtain a formally verified
SAT solving algorithm. When run on a CNF formula it produces a model
satisfying the formula or a DPLL derivation showing its
unsatisfiability.
We also prove the equivalence of DPLL and resolution and extract
a program translating DPLL proofs into resolution proofs of smaller 
or equal size.

Minlog \cite{MinlogSystem,HB98,UB11} is an interactive proof assistant
based on a first-order natural deduction calculus. 
It implements various methods of program extraction such as realizability \cite{KR59}
(which can be viewed as a technical rendering of the Curry-Howard
correspondence \cite{HC58,WH80}) and the Dialectica interpretation. 
It also extends program extraction to classical proofs via the Friedman/Dragalin
$A$-translation. All these techniques are refined and optimized in order to
improve usability and to obtain simpler programs. In addition to
extracting a program from a proof, Minlog also automatically extracts
a proof that the program meets its specification; see for instance
\cite{SW11} for an overview on program extraction and its underlying
theory.  A number of substantial case studies on program extraction
have been carried out reaching from the extraction of a
normalization-by-evaluation algorithm~\cite{UB06} to the extraction of
programs in constructive analysis~\cite{Schwichtenberg08}.  Recent
developments concentrate on program extraction for induction and
coinduction, including applications in the context of exact real
number computation~\cite{UB12}.

An optimization in Minlog that is particularly important for this
paper is the use of so-called non-computational quantifiers, which
flag certain information in the proof as computationally irrelevant,
and therefore allow for the removal of computational redundancy in the
extracted program. In case of the extracted SAT solver, this leads to
a significant
improvement. 
%
%

We also applied an automatic translation of Minlog terms into Haskell
code to the extracted program and observed a further dramatic
improvement of performance.  We evaluate the performance of our
extracted solver by comparing it 1) with another verified SAT solver,
Versat~\cite{DO12}, using Pigeon hole formulae and 2) with an
industrial tool, SCADE~\cite{SCADE}, by means of an example from the railway
domain.


An earlier version of this article, containing partial results, 
was reported at the MFPS~2012 \cite{LBS12} conference.

\subsection{Related Work}

There are several other systems supporting program extraction from
proofs for the purpose of producing formally verified programs.  An
early example is the Nuprl system \cite{RC86}; other mature
interactive theorem provers that implement program extraction are Coq
\cite{YB04}, which is based on the Calculus of Inductive
Constructions, and Isabelle \cite{TN99}, a generic theorem prover with
extensions for many logics (see \cite{BN02} for code generation and
\cite{Berghofer03} for program extraction from proofs in Isabelle).
More recently, other interactive theorem provers based on dependent
types \cite{PM80}, such as Agda~\cite{AB09} and Idris~\cite{Brady13},
have emerged which realize the Curry-Howard correspondence and
therefore can also be viewed as supporting program extraction.



The Coq system has been used in several approaches to formalize
automatic theorem proving. 
Lescuyer and Conchon~\cite{SL08} program a SAT solver based on the DPLL
algorithm as a recursive function in Coq, and verify its soundness and completeness formally in the system.
The solver is then instantiated on the
propositional fragment of Coq's logic, creating a user friendly proof
tactic.  Similarly, Verma et al.~\cite{KV00} formalize Binary Decision Diagrams
in Coq, prove their correctness, and extract certified BDD
algorithms in OCaml.
The main reason for their formalization was to integrate symbolic
model checking in Coq.
Significant work has also been performed in Isabelle with several
decision procedures verified and integrated into the
system. The DPLL algorithm has been formalized by Mari\'{c} and
Jani\v{c}i\'{c}~\cite{FM10}.  This approach was extended to formalize
a SAT solver including optimizations such as clause learning and the
lazy two-watched-literal data structure \cite{FM10b}.  The authors
investigated automatic code generation, but in the end the verified
algorithm was manually translated into C code.  The automatic theorem
prover Metis \cite{LP07} 
is used inside Isabelle to reconstruct proofs from faster external
procedures such as the ones used in Sledgehammer \cite{SB10}.  A
different direction to deal with the correctness of SAT solvers has
been to verify a proof checker for resolution proofs \cite{NW13}. This
will check and guarantee that the output from a solver for a
particular SAT problem is correct.

The DPLL solver Versat~\cite{DO12}, mentioned earlier,
was formalized and verified in the dependently typed programming
language Guru \cite{AS09} and then translated into imperative C
code. This translation is possible because Guru contains mutable
arrays. Since Guru allows for the verification of low level
optimizations involving such arrays and Versat implements
clause learning, the resulting solver is quite efficient. However,
this approach differs from ours in that only soundness has been proven
for Versat, whilst we have the possibility to deliver a proof in the
case of unsatisfiability. This means that while every satisfiable
assignment produced by Versat can be trusted, it is not guaranteed
that Versat can solve every solvable problem.

A program extraction project related to ours was carried out by
Weich~\cite{Weich98} who gave two constructive proofs of the decidability of
intuitionistic propositional logic and extracted two different programs that,
for a given formula, either produce a derivation in intuitionistic
sequent calculus, or a Kripke counter-model. The second proof and program
extraction were formalized in Minlog for the implicational fragment.
 
The articles \cite{SL08,FM10} verifying a DPLL SAT solver (in
both Coq and Isabelle) were the main motivation for our work.
Their approaches involve a formalization of the algorithm to be
verified. In contrast, we work in a system that does not require any 
formalization of algorithms. It is enough to prove that each CNF-formula 
is either unsatisfiable or has a model. The desired SAT solving 
algorithm and its correctness proof are then extracted fully automatically.
%



\section{Preliminaries}

We begin with some basic definitions, following \cite{SL08,FM10}.

\hspace{1mm}

\begin{defi}
\mbox{}
\begin{enumerate}
\item A \emph{literal} $l$ is either a positive variable $+v$ or a negative variable $-v$, i.e.\ a variable $v$ with a label $+$ or ${-}$ attached.

\item For every literal $l$ we define the opposite literal $\mybar{l}$ by $\mybar{+v}= -v$, $\mybar{-v} = +v$. 

\item We set $\var(+v) = \var(-v) = v$, $\var(L) = \{\var(l) \mid l\in L\}$
for a set of literals $L$, and 
$\var(\Delta) = \bigcup\{\var(L)\mid L\in\Delta\}$ for a set of sets of 
literals $\Delta$.

\item A \emph{clause} $C$ is a finite set of literals 
to be viewed as their disjunction.

\item A formula in \emph{conjunctive normal form} (CNF) is a 
finite conjunction of clauses. 
%
By a \emph{formula} $\Delta$ we will always mean a formula in CNF,
and we will identify it with a finite set of clauses 
$\{ C_1 , \ldots , C_k \}$, representing the conjunction of the $C_i$.

\item A \emph{valuation} $\Gamma$ is a finite set of literals 
to be viewed as their conjunction.

\item A valuation $\Gamma$ is \emph{consistent} if 
$\forall l \,( l \in \Gamma \to \mybar{l} \notin \Gamma)$.
We let $\consvals$ denote the set of all consistent valuations.

\item A \emph{model} is a total function $M$ which maps literals\footnote{We map literals instead of variables as a model is constructed from a set of literals in the form of a valuation.} to booleans and satisfies the property
$\forall l \, (M \ l \leftrightarrow \neg M \ \mybar{l})$.
\end{enumerate}
We shall use the abbreviations 
\begin{itemize}
\item $M \models \Gamma$, for $\forall l \in \Gamma \, (M \ l)$ 
(`$M$ is a model of $\Gamma$'),
\item $M \models \Delta$, for 
$\forall C \in \Delta \, \exists l \in C \,(M \ l)$ 
(`$M$ is a model of $\Delta$').
\end{itemize}
We call a valuation $\Gamma$ and a formula $\Delta$ \emph{compatible} 
if there exists a model satisfying both, i.e. 
$\exists M \, (M \models \Gamma \wedge M\models \Delta)$;
otherwise $\Gamma$ and $\Delta$ are called \emph{incompatible}. 
\end{defi}
%
A \emph{sequent} $\Gamma \vdash \Delta$ is a pair consisting of a valuation and a formula.
%
The intended meaning of a sequent $\Gamma \vdash \Delta$ is that $\Gamma$ and $\Delta$
are incompatible. As a special case, when $\Gamma$ is empty, $\vdash \Delta$ means that $\Delta$ is unsatisfiable. 
%
In the following we use the notations $X,a := \{x \mid x\in X \lor x = a\}$ 
and 
$X\setminus a := \{x \mid x\in X \land x \neq a\}$. 
\begin{defi}[DPLL Proof System] The DPLL proof system consists 
of five rules: 
\label{def:proofsystem-DPLL}
\bigskip \\
\begin{center}
\AxiomC{$\Gamma, l \vdash \Delta $}
\RightLabel{($\Unit$)}
\UnaryInfC{$\Gamma \vdash \Delta, \{l \} $}
\DisplayProof \
\qquad
\AxiomC{$\Gamma, l  \vdash \Delta, C$}
\RightLabel{($\Red$)}
\UnaryInfC{$\Gamma, l \vdash \Delta, (C,\mybar{l})$}
\DisplayProof \
\qquad
\AxiomC{$\Gamma, l \vdash \Delta$}
\RightLabel{($\Elim$)}
\UnaryInfC{$\Gamma, l \vdash \Delta,(C,l)$}
\DisplayProof \

\bigskip

\AxiomC{$$}
\RightLabel{($\Conflict$)}
\UnaryInfC{$\Gamma \vdash \Delta,  \emptyset$}
\DisplayProof \
\qquad
\AxiomC{$\Gamma,l \vdash \Delta$}
\AxiomC{$\Gamma, \mybar{l} \vdash \Delta$}
\RightLabel{($\Split$)}
\BinaryInfC{$\Gamma  \vdash \Delta$}
\DisplayProof \
\end{center}
\end{defi} 

Several variants of the DPLL proof system have featured in the literature. The above definition is closest to the Coq formalisation \cite{SL08}, other formalisations such as \cite{FM10} and \cite{JH09} combine the $\Unit$, $\Red$ and $\Elim$ rules to form a single rule called the "1-literal rule" or "unit propagation".


\section{Soundness and Completeness}

\subsection{Soundness and Completeness of DPLL}

In this section we sketch the formal proof of soundness and completeness
of the DPLL proof system. We will be very brief with the Soundness Theorem
since its proof does not carry computational content and a similar proof is carried out in~\cite{SL08,FM10}. On the other hand,
we will describe the proof of the Completeness Theorem in some detail since
we extract our SAT solver from it.

We first reformulate the DPLL proof system as an inductive definition
that can be immediately formalized in the Minlog system. The definition
has a clause for each rule. We notationally identify a sequent
$\Gamma \vdash \Delta$ with the statement `$\Gamma \vdash \Delta$ is derivable'.

\pagebreak 
\begin{rem}
\label{def:derivable-DPLL}
The proof system described in Definition \ref{def:proofsystem-DPLL} has been reformulated for our theorem prover. The set of sequents  $\Gamma \vdash \Delta$ is 
defined inductively by the following (universally quantified) inductive clauses:
\begin{center}
\begin{tabular}{ll}
$\Conflict$ &
$\emptyset \in \Delta \to \Gamma \vdash \Delta$
\\
$\Unit$ &
$\{ l \} \in \Delta \to \Gamma, l \vdash 
\Delta \setminus \{l \} \to  \Gamma \vdash  \Delta$
\\
$\Elim$ &
$l \in \Gamma \to l \in  C \to  C \in \Delta \to  
\Gamma \vdash  \Delta \setminus C  \to   \Gamma \vdash \Delta $
\\
$\Red$ &
$l \in \Gamma \to \mybar{l} \in C \to C \in \Delta \to \Gamma \vdash  
(\Delta \setminus C) , (C \setminus \mybar{l}) \to   \Gamma \vdash \Delta$
\\
$\Split$ &
$\Gamma, l  \vdash \Delta 
\to  \ \Gamma, \mybar{l} \vdash  \Delta \to \Gamma \vdash \Delta$
\end{tabular}
\end{center}
\end{rem}

\begin{thm}[Soundness]
If $\Gamma \vdash \Delta$, 
then $\Gamma$ and $\Delta$ are incompatible.

\end{thm}
The proof proceeds by structural induction on the given derivation 
of the sequent $\Gamma \vdash \Delta$. We omit further details.
%

We now turn our attention to the Completeness Theorem for the DPLL proof 
system. The expected statement of completeness is:
$$ \forall \Gamma\in\consvals, \forall \Delta\, 
      (\incompatible{\Gamma}{\Delta} \to \Gamma \vdash \Delta). $$

A constructive proof of this statement would yield a program that
computes a DPLL proof for incompatible $\Gamma$, $\Delta$.
We reformulate the statement by replacing the implication
`$\incompatible{\Gamma}{\Delta} \to \Gamma \vdash\Delta$' with
the classically equivalent but constructively
stronger disjunction  
`$\compatible{\Gamma}{\Delta} \vee \Gamma \vdash \Delta$'.
In this way, we obtain an enhanced program that still computes a DPLL
proof for incompatible $\Gamma$, $\Delta$, but in addition produces a model
if $\Gamma$ and $\Delta$ are compatible.

\begin{thm}[Completeness of DPLL] 
\label{thm:dpllcompleteness}
$$ \forall \Gamma\in\consvals, \forall \Delta\,
     (\compatible{\Gamma}{\Delta}  \lor  \Gamma \vdash \Delta) $$

\proof
{\rm
We aim to perform the proof in such a way that an efficient program 
is extracted. Therefore, we adopt the following strategy:
\begin{enumerate}
 \item Since performing a $\Split$ rule is the only computational expensive 
    operation
     -- it is the only rule forcing the proof search to branch -- we only
    apply it if absolutely necessary.

 \item  We perform an optimization on the proof level by partitioning the clauses
    into `clean' and `unclean' clauses, where a clause is called clean if we
    cannot apply $\Elim$, $\Red$ or $\Unit$ to that clause.
    This increases the efficiency of the algorithm by reducing the number
    of comparisons needed.
\end{enumerate}
To this end we show that for all valuations $\Gamma$, and formulae $\Delta$, $\Theta$,\\[1em]
\hspace*{3em}$\emptyset \notin \Theta \wedge 
\Gamma\in\consvals \land \var(\Gamma) \cap \var(\Theta) = \emptyset\to$\\[.5em]
\hspace*{12em}$(\Gamma \vdash  \Delta\cup\Theta) \lor 
 \exists M(M \models \Gamma \land M \models \Delta\cup\Theta)$.\\[1em]
The proof is by main induction on the measure
$$\muu{\Gamma}{\Delta}{\Theta} := \measure{\Gamma}{(\Delta\cup\Theta)}+
\weight{\Delta}+\weight{\Theta}$$
where
\begin{center}
\begin{tabular}{lll}
$|X|$             &$:=$& \hbox{the cardinality of}\,$X$\\
$\Delta \setminus\!\!\setminus V $
                &$:=$& $\{ l| \exists C\in \Delta(l \in C) \land \var (l) \notin V \}$\\
$\weight{\Delta} $&$:=$&$ \sum_{C\in\Delta}|C|$ 
\end{tabular}
\end{center}
and a side induction on $|\Delta|$ (i.e.~the number of clauses in $\Delta$).
\par

\bigskip

Let $\Gamma$, $\Delta$, $\Theta$ be given such that
$\emptyset \notin \Theta$, $\Gamma\in\consvals$, and 
$\var(\Gamma) \cap \var(\Theta) = \emptyset$. \\[1em]
\noindent\emph{Case 1 $\Delta = \emptyset$.} \\[1em]
\noindent\emph{Case 1.1 $\Theta = \emptyset$.}\\ 
We define a model $M$ by 
$M(l) = \True \leftrightarrow l \in \Gamma$. Then 
$M \models \Gamma \land M \models \emptyset$ holds.\\[1em]
\noindent\emph{Case 1.2 $\Theta \neq \emptyset$.}\\  
Let $C$ be a clause in $\Theta$ and let $l\in C$ ($C\neq\emptyset$, by the 
assumption on $\Theta$). Then 
$\muu{(\Gamma,l)}{\Theta}{\emptyset} < \muu{\Gamma}{\emptyset}{\Theta}$
since $\measure{\Gamma,l}{\Theta} < \measure{\Gamma}{\Theta}$ and $\weight{\Theta} + \weight{\emptyset} = \weight{\emptyset} + \weight{\Theta}$. 
Furthermore, for the values $(\Gamma,l)$, $\Theta$, $\emptyset$
the hypotheses of the theorem are clearly satisfied.  
Hence the induction hypothesis for these values yields
\begin{equation}\label{eq-split-left}
(\Gamma,l \vdash  \Theta) \lor 
 \exists M(M \models \Gamma,l \wedge M \models \Theta)
\end{equation}
Similarly, we can apply the induction hypothesis to 
$(\Gamma,\mybar{l})$, $\Theta$, and $\emptyset$ yielding
\begin{equation}\label{eq-split-right}
(\Gamma,\mybar{l} \vdash  \Theta) \lor 
 \exists M(M \models \Gamma,\mybar{l} \wedge M \models \Theta)
\end{equation}
The disjunctions \eqref{eq-split-left} and \eqref{eq-split-right} result
in 4 cases:
In the case that $\Gamma,l \vdash \Theta$ and $\Gamma, \mybar{l} \vdash \Theta$ hold 
the $\Split$ rule is applied and we obtain $\Gamma \vdash \Theta$. 
In all other cases we use one of the models obtained from the 
induction hypotheses. \\[1em]
\noindent\emph{Case 2 $\Delta = \Delta', C$.}\\
We perform a case distinction on whether the valuation $\Gamma$ has a 
literal  in common with $C$.\\[1em]
\noindent\emph{Case 2.1  $ \Gamma \cap C = \emptyset$.}\\
We perform a further case distinction on the cardinality of the clause $C$.\\[1em]
\noindent\emph{Case 2.1.1  $C = \emptyset$.}\\
It suffices to show $\Gamma \vdash (\Delta' , \emptyset) \cup \Theta$. 
This follows from the $\Conflict$ rule.\\[1em]
\noindent\emph{Case 2.1.2 $C = \{ l \}$}.\\
If $\mybar{l}\in\Gamma$, then $\Gamma \vdash (\Delta' , \{ l \}) \cup \Theta$
can be derived by applying (in backwards fashion) the $\Red$ rule followed
by the $\Conflict$ rule.
If $\mybar{l}\notin\Gamma$, then we use the induction
hypothesis with $(\Gamma,l) $, $\Delta' \cup \Theta$, $\emptyset$.
This is possible since $\muu{(\Gamma,l)}{\Delta' \cup \Theta}{\emptyset}
<\muu{\Gamma}{(\Delta',\{l\})}{\Theta}$ because 
$ \measure{\Gamma,l}{(\Delta'\cup \Theta)} < \measure{\Gamma}{(\Delta'\cup(\{l\},\Theta))}
$ and 
$\weight{\Delta'\cup\Theta} < \weight{\Delta',\{l\}}+\weight{\Theta}$.
Since for the values $(\Gamma,l) $, $\Delta' \cup \Theta$, $\emptyset$
the hypotheses of the theorem are satisfied (i.p.\ $\Gamma,l$ is consistent since
$\mybar{l}\notin\Gamma$), we obtain the disjunction 
$(\Gamma, l \vdash \Delta' \cup
\Theta) \vee \exists M(M \models \Gamma, l \wedge M \models
(\Delta' \cup \Theta))$. 
In the case that $\Gamma,l \vdash \Delta'
\cup \Theta$ holds we apply the $\Unit$ rule resulting in $\Gamma
\vdash \Delta \cup \Theta$. 
In the other case we have a model of  $\Gamma, l$ and $\Delta' \cup \Theta$
which clearly also models $\Gamma$ and $\Delta \cup \Theta$.\\[1em]
\noindent
\emph{Case 2.1.3 $|C| \geq 2$}.\\
We perform a case distinction on $\exists
l \,(l \in C \wedge \mybar{l} \in \Gamma) \lor \neg \exists l (l \in C
\wedge \mybar{l} \in \Gamma)$. This disjunction can be proven constructively, 
since the sets involved are finite.\\[.5em]
\noindent
\emph{Case 2.1.3.1 $\mybar{l} \in \Gamma$ for some $l\in C$}.\\
Then we have $\muu{\Gamma}{(\Delta',C\setminus l)}{\Theta}
<\muu{\Gamma}{(\Delta',C)}{\Theta}$ since 
$\weight{\Delta',C\setminus l}<\weight{\Delta',C}$ and $\measure{\Gamma}{(\Delta',C\setminus l)} = \measure{\Gamma}{(\Delta',C)}$ .
The hypotheses of the theorem are satisfied for the chosen values.
Hence we obtain, by induction hypothesis,
$(\Gamma \vdash (\Delta', (C \setminus l)) \cup \Theta)\lor 
\exists M(M\models\Gamma \land M\models(\Delta',(C\setminus l))\cup\Theta)$. 
In the case that $\Gamma \vdash (\Delta',(C \setminus l)) \cup \Theta$ holds, 
we apply the $\Red$ rule. 
In the other case we have a model of $\Gamma$ and 
$(\Delta', (C\setminus l)) \cup \Theta$ which also models the weaker formula
$(\Delta',C) \cup \Theta$.\\[.5em]
\noindent
\emph{Case 2.1.3.2} $\neg \exists l\, (l \in C \wedge \mybar{l} \in
\Gamma)$.\\
In this case we may move $C$ from $\Delta$ to $\Theta$:
Since $\muu{\Gamma}{\Delta'}{(\Theta,C)}
\le\muu{\Gamma}{(\Delta',C)}{\Theta}$ we can apply the 
side induction hypothesis
to $\Gamma$, $\Delta'$, $(\Theta,C)$. Since for these values the hypotheses
of the theorem are satisfied we obtain
$\Gamma \vdash \Delta' \cup (\Theta, C) \lor
\exists M(M \models \Gamma \land M \models \Delta' \cup (\Theta,C))$
which is the same as the required disjunction
$\Gamma \vdash (\Delta',C) \cup \Theta \lor
\exists M(M \models \Gamma \land M \models (\Delta',C) \cup \Theta)$. \\[1em]
\noindent\emph{Case 2.2  $\Gamma \cap C \neq \emptyset$.}\\
We can prove constructively that in this case 
$\Gamma$ and $C$ have some literal $l$ in common.
We apply the induction hypothesis to 
$\Gamma$, $(\Delta' ,(C \setminus l))$, $\Theta$. Since clearly the measure 
decreases, $\weight{\Delta',(C \setminus l)}<\weight{\Delta',C}$ and 
$\measure{\Gamma}{(\Delta' ,(C \setminus l))} = \measure{\Gamma}{(\Delta',C)}$, 
and the hypotheses of the theorem are satisfied, we obtain
$\Gamma \vdash (\Delta',(C \setminus l)) \cup \Theta$ or 
$\exists M(M\models\Gamma \land M\models(\Delta',(C \setminus l))\cup\Theta)$. 
In the first case we apply the $\Elim$ rule, in the second case we use the model
provided.
}
\qed

\end{thm}


\subsection{Resolution}

The resolution proof system~\cite{AR65} is widely used in practical
applications, for instance in tools for proof checking and debugging
\cite{tracecheck} or interchange between different solvers
\cite{CK13}. State-of-the-art SAT solvers such as MiniSAT
\cite{miniSAT} and zChaff \cite{MM01} return (extended) resolution
proofs for unsatisfiable problems.  By formalizing that every DPLL
derivation has an equivalent resolution derivation, and combining this result
with the completeness proof from the previous section, we can extract
a SAT solver which produces resolution derivations. The equivalence of DPLL
and resolution was first shown by Robinson \cite{JR68} who translated
between the two proof systems using semantic trees.


By enriching the systems with size information we are able to 
show that the size of the resulting resolution proof does not exceed
the size of the original DPLL proof.


For every valuation $\Gamma$ we define a clause $\mybar{\Gamma}$
representing its negation by $ \mybar{ \{l_1, \ldots , l_k\}} =
\{\mybar{l_1}, \ldots ,\mybar{l_k} \}$.

\begin{defi}[Resolution Proof System] The derivable resolution sequents $\Gamma \modres{n} C$ with a derivation of size $n$ are conveniently defined by two rules: subsumption (or axiom) and resolution.
\bigskip

\label{def:resolutionps}
\begin{center}
\AxiomC{\phantom{$\Delta \modres{n} C \vee l$}}
\RightLabel{($\Sub$)  $C \subseteq C'$ }
\UnaryInfC{$\Delta,C \modres{0} C'$}
\DisplayProof
\qquad
\AxiomC{$\Delta \modres{n} C \vee l$}
\AxiomC{$ \Delta \modres{m} C' \vee \bar{l}$}
\RightLabel{($\Res$)}
\BinaryInfC{$\Delta \modres{n + m + 1} C \vee C'$}
\DisplayProof 
\end{center}
\bigskip
\end{defi}

We also need a version of the DPLL proof system with added bounds in order to speak about the sizes of the proofs.

\begin{rem}[Derivable refined DPLL sequents]
  The derivable DPLL sequents $\Gamma \moddpll{n} \Delta$ with a
  derivation of size $n$ are inductively defined by the following
  clauses:
\begin{center}
\begin{tabular}{ll}
$\Conflict$ &
$\emptyset \in \Delta \to \Gamma \moddpll{0} \Delta$
\\
$\Unit$ &
$\{ l \} \in \Delta \to \Gamma, l \moddpll{n} 
\Delta \setminus \{l \} \to  \Gamma \moddpll{n + 1}  \Delta$
\\
$\Elim$ &
$l \in \Gamma \to l \in  C \to  C \in \Delta \to  
\Gamma \moddpll{n}  \Delta \setminus C  \to   \Gamma \moddpll{n + 1} \Delta $
\\
$\Red$ &
$l \in \Gamma \to \mybar{l} \in C \to C \in \Delta \to \Gamma \moddpll{n}  
(\Delta \setminus C) , (C \setminus \mybar{l}) \to   \Gamma \moddpll{n + 1} \Delta$
\\
$\Split$ &
$\Gamma, l  \moddpll{n} \Delta 
\to  \ \Gamma, \mybar{l} \moddpll{m}  \Delta \to \Gamma \moddpll{n + m + 1} \Delta$
\end{tabular}
\end{center}
\label{def:derivable-DPLL-refined}
\end{rem}

\begin{rem}
  The resolution proof system from Definition \ref{def:resolutionps}
  has been reformulated as follows for our theorem prover. The
  derivable resolution sequents $\Gamma \modres{n} C$ with a
  derivation of size $n$ are inductively defined by the following
  clauses:
\begin{center}
\begin{tabular}{ll}
$\Sub$ &
$C_0 \in \Delta \to C_0 \subseteq C \to \Gamma \modres{0} C$
\\
$\Res$ &
$\Delta \modres{n} (C' \vee \mybar{l}) 
\to  \ \Delta \modres{m} (C \vee l)  \to \Delta \modres{n + m + 1} (C \vee C')$
\end{tabular}
\end{center}
\label{def:resolution}
\end{rem}




\begin{thm}[DPLL implies Resolution] 
\label{thm:dpllresolution}
For all consistent valuations $\Gamma$, CNF formulae $\Delta$ and natural numbers $n$: If $\Gamma \moddpll{n} \Delta$, then $\Delta \modres{m} \bar{\Gamma}$
for some $m\le n$.
\proof
The proof is an easy induction on DPLL derivations. 
We only sketch the overall idea. 
The $\Conflict$ and $\Split$ rule translate into
the $\Sub$ and $\Res$ rule respectively.
Both of these rules have the same cost to perform them as the DPLL rules and so the size of the derivations are less or equal. 
An application of the $\Unit$ rule is a special case of the $\Res$ rule in which one of the branches is obtained via a subsumption of a unit clause. The size of these two proofs is less or equal since the cost of performing the $\Sub$ rule and $\Res$ rule together is the same as that of the $\Unit$ rule. Finally, both the $\Elim$ and $\Red$ DPLL rules correspond to a form of weakening in the resolution proof which is done at no cost because the resulting resolution proofs 
are smaller in size than the DPLL proofs. 
\qed

\end{thm}





\begin{rem}
One can also easily prove that resolution implies DPLL, more precisely, if
$\Delta \modres{} C$, then  $\bar{C} \moddpll{} \Delta$.
However, as long as the sizes of derivations are measured only in terms of
the number of applications of rules (as we do above), no size bound can be given.
The reason is that the translation of one instance of the subsumption rule
\bigskip

\begin{center}
\AxiomC{$ $}
\RightLabel{($Sub$)  $C \subseteq C'$ }
\UnaryInfC{$\Delta,C \modres{0} C'$}
\DisplayProof
\end{center}
into DPLL requires $n$ applications of the $\mathbf{Red}$ rule where $n$ is the
number of literals in $C$.
\end{rem}
\bigskip 

The Completeness Theorem for DPLL (Theorem~\ref{thm:dpllcompleteness}), 
adapted to the DPLL system with size information, and 
Theorem~\ref{thm:dpllresolution}~(a) immediately imply:
\begin{thm}[Completeness of the Resolution Proof System]
\label{thm:resolutioncomplete}

$$ \forall \Delta \,  ((\exists M \, M \models \Delta) \vee (\exists n \, \Delta \modres{n} \emptyset))$$


\end{thm}

The program extracted from Theorem~\ref{thm:dpllresolution} translates 
DPLL derivations into equivalent resolution derivations.
This translator and the SAT solver extracted from the Completeness Theorem for 
DPLL (Theorem~\ref{thm:dpllcompleteness}) are combined in the
program extracted from Theorem~\ref{thm:resolutioncomplete}
to a SAT solver that yields resolution refutations for unsatisfiable formulae.
Since the computationally hard and interesting part of this program is 
entirely contained in the DPLL-based SAT solver, we will restrict
our attention to the latter when we discuss the extracted programs in detail 
in Sect.~\ref{sec:program}.




\section{Program Extraction}
%
\subsection{Theory}
Program extraction in Minlog is based on modified realizability
\cite{KR59}.  We highlight a few aspects that are important to
understand the optimizations we achieved.  For a complete and precise
description of program extraction we refer to \cite{SW11}.
A formula is said to have \emph{computational content} if it has at 
least one occurrence of $\exists$ or $\lor$ at a strictly positive position.
To every such formula $A$ one assigns a type $\tau(A)$ 
of 'potential realizers'. If the formula has no computational content,
one sets $\tau(A)=\epsilon$. From a proof of a formula $A$ 
with computational content one can extract a program $M$ of type 
$\tau(A)$ that \emph{realizes} $A$ (written $\ire{M}{A}$), 
that is, $M$ solves the computational problem expressed by $A$.
In order to fine-tune the computational content, in particular to remove
redundant content, Minlog offers, besides the usual quantifiers $\forall$ and
$\exists$, the \emph{non-computational (nc)} quantifiers $\forallnc$ and 
$\existsnc$ (which roughly correspond to quantification in Prop in Coq).
These have the same logical meaning as the usual quantifiers, but indicate that the extracted program does not
operate on the quantified variable, only on its realizer.
%
%
%
The definitions of the type and the realizability relations for the 
ordinary universal quantifier contrasted with its $\NC$ version are:

\[
\begin{array}{lclclcl}
   \tau(\forall x^\rho A)   &=& \rho\to\tau(A) &\qquad& 
      \ire{f}{\forall x^\rho A} &=& \forall x^\rho (\ire{f(x)}{A})\\
   \tau(\forallnc x^\rho A) &=& \tau(A) &&
         \ire{a}{\forallnc x^\rho A} &=& \forall x^\rho (\ire{a}{A})
\end{array}
\]
Similarly for the two versions of the existential quantifier:
\[
\begin{array}{lclclcl}
   \tau(\exists x^\rho A) = \rho \times \tau(A) &\qquad&
        \ire{(a,y)}{\exists x^\rho} A = \ire{a}{A[y/x]}\\
   \tau(\existsnc x^\rho A) = \tau(A)  &&
           \ire{a}{\existsnc x^\rho A} = \exists x^\rho (\ire{a}{A})
\end{array}
\]
One sees that for the $\NC$-quantifiers the realizers do not depend on the quantified variables.
The program extraction procedure respects the different kind
of quantifiers by omitting in the $\NC$ case any information corresponding
to the quantified variable. The proof rules for the $\NC$-quantifiers are
subject to
stricter variable conditions ensuring that the omitted information
is indeed not needed in the extracted program. Minlog is able to
automatically detect the maximal set of occurrences of quantifiers in a proof
that can be made non-computational without compromising the correctness
of the proof~\cite{DR09}. This holds for the logical parts of the proof only. 
In the formalization of inductive definitions one has to manually place $\forall_{nc}$ quantifiers.

\subsection{Extraction to Haskell}

The programs extracted by Minlog are terms in Minlog's internal term
language. This has the advantage that extracted programs can be reused
for further proofs, and properties of the programs can be formally
proven, again inside Minlog. Furthermore, the extracted programs are
provably correct, and a (soundness) proof of this fact is
automatically generated by Minlog. However, there are also inherent
disadvantages: the interoperability of the extracted programs with
external libraries or devices is limited, and executing the programs
is sometimes slow.
For both these reasons, it makes sense to translate the extracted
programs into more conventional, general-purpose programming
languages. Minlog implements a translation to Haskell (and also a
limited translation to Scheme). Extracting to a lazy language such as
Haskell makes the treatment of coinduction and corecursion
(which is not used in our example)
particularly simple~\cite{MNFS13}.

There is a close fit between Haskell and the Minlog term language, and
the translation is quite straightforward; basic terms such as
variables, lambda abstractions, etc are translated to the
corresponding Haskell terms. Standard algebras such as e.g.\ lists,
integers, booleans, sum and product types are translated to their
implementation in the Haskell Prelude, while user-defined algebras in
general are translated to algebraic data types. Natural numbers are
translated to (unbounded) integers for efficiency.\footnote{Using
  bounded \texttt{Int}s instead of unbounded \texttt{Integer}s would
  of course not be sound. In the cases where it would be safe to do so, it would also not result in any particular performance gains, since GHC stores small \texttt{Integer}s as \texttt{Int}s.}
Program constants and their computation rules in Minlog correspond to
functions defined by pattern matching in Haskell. Some care must be
taken for e.g.\ the natural numbers; in Minlog, pattern matching on
natural numbers is possible, but natural numbers are translated to
integers, for which no pattern matching is available in
Haskell. Instead guard conditions have to be used.
Recursion operators, realizing structural induction, are automatically
generated as Haskell functions by the translation. Minlog also
supports general recursion along a decreasing measure, which makes
sure that the program terminates. The Minlog implementation of the
general recursion operator ensures that recursive calls are only made
on arguments that are smaller than the current argument with respect
to the measure:
\begin{align*}
&\grec : (\rho\to\Nat)\to\rho\to(\rho\to(\rho\to\tau)\to\tau)\to\tau\\
&\grec(\mu, x, f) = f(x,(\lambda y\,.\,
   \Boolelim{\mu(y) < \mu(x)}{\grec(\mu, y, f)}{\inhab{\tau}}))
\end{align*}
($\inhab{\tau}$ is a canonical inhabitant of $\tau$, justified by
the fact that all domains are inhabited in the intended, standard
semantics).
Note that the (potentially expensive) test $\mu(y) < \mu(x)$ 
is computationally unnecessary, since at runtime we already know 
that our extracted program will only use recursive calls on smaller 
arguments. However, this test is needed because of Minlog's eager 
evaluation strategy. Omitting the test:
\begin{align}
&\grec(x, f) = f(x, (\lambda y\,.\,\grec(y, f))) \label{eq:grec-eff}
\end{align}
would make Minlog get stuck in an endless loop, forever evaluating the
recursive call $\grec(y, f)$ regardless of whether it is going to
be used or not. 

However, since Haskell is a lazy language, we can safely implement
general recursion using \eqref{eq:grec-eff}. This can give large
efficiency gains in certain situations (see
Section~\ref{sec:compare-performance}). In a lazy setting, soundness
of this variant of the program extraction process can still be proven,
and the Haskell translation supports this optimization. However, there
is now a discrepancy between Minlog programs and their Haskell
translations: if called in a way that does not respect the measure,
the Minlog implementation of $\grec$ will halt with an arbitrary
value, while the Haskell version will diverge. For this reason, the
optimization can be turned on and off with a switch, if identical
behavior is important. Of course, every extracted term will respect
the measure.

\section{The Extracted Program}

\label{sec:program}
The size of the DPLL formalization 
is approximately 5500  lines of Minlog code. The extracted program comes to 300 lines of code as a Minlog term and  600 lines of Haskell code. In the following we present two versions of our extracted solver: one optimized with $\forallnc$ quantifiers which we shall refer to as the $\forallnc$ solver, and the other without these optimizations which we shall refer to as the $\forall$ solver.
  
%
%
The $\forall$ solver takes a CNF formula $\Delta$ represented as a list of
clauses as input, and produces either a model of $\Delta$ or a
derivation of its unsatisfiability.
%
%
Models are represented as functions from literals to booleans. An
algebraic data type for DPLL derivations is automatically generated
from its inductive definition in Minlog. It has five constructors, one
for each of the DPLL rules in Definition~\ref{def:derivable-DPLL}:
\begin{verbatim}
      data Algdpll = CConflict Valu For
                   | CElim Valu For Cla Lit Algdpll
                   | CUnit Valu For Lit Algdpll
                   | CRed Valu For Cla Lit Algdpll
                   | CSplit Valu For Lit Algdpll Algdpll
        deriving (Show, Read, Eq, Ord)
\end{verbatim}
Each constructor takes a formula and a valuation as arguments. The
formula itself never changes during the proof and is only part of the algebra for the purpose of proving correctness and does not play a role in any computation. While the valuation changes
during the proof search, these changes can be captured by indicating
which literal was added by the $\Unit$ and $\Split$ rules, thus making
the valuation redundant as well.  We added $\NC$-quantifiers to the
definition by hand in order to remove redundant
computational content, resulting in
\begin{verbatim}
      data Algdpll = CConflict
                   | CElim Cla Lit Algdpll
                   | CUnit Lit Algdpll
                   | CRed Cla Lit Algdpll
                   | CSplit Lit Algdpll Algdpll
        deriving (Show, Read, Eq, Ord)
\end{verbatim}
%
%

The control structure of the program closely follows the structure
of the case distinctions and proofs by induction performed in the
proof. Lemmas invoked during the proof are extracted separately and
called as procedures. Since the proof is by general induction along a
measure, the main body of the program is using general recursion along
the same measure.

\section{Execution of the Extracted Program}

%
%
In the following we will see how both $\forall$ and $\forallnc$ solvers 
behave when they are applied to a number of SAT problems. The extracted 
decision procedure was run on several instances of the pigeon hole 
principle \cite{SC79} in both Minlog and as Haskell programs. The pigeon hole principle states that there is 
no injective function that maps 
$\{ 1, 2 \ldots , n \}$ to $\{1,2, \ldots, n-1 \}$.
%

\begin{defi}[Pigeon Hole Formula]
$\mathbf{PHP}(n,m) := \{\{l_{i,1},\ldots,l_{i,m}\} | 1 \leq i \leq n  \} \cup \{ \{\mybar {l_{i,k}},\mybar{ l_{j,k}}\} | 1 \leq i < j \leq n, 1 \leq k \leq m \}  $
 \end{defi}

Here $l_{i,k}$ represents the statement ``pigeon $i$ sits in hole $k$''.
The whole formula $\mathbf{PHP}(n,m)$ states that $n$ pigeons sit in $m$ holes such that no two pigeons are in the same hole.
Hence, $\mathbf{PHP}(n,m)$ is satisfiable iff $n \leq m$. 
For example, if we run our DPLL solver with the formula
$\mathbf{PHP}(2,1) = \{\{l_{11} \}, \{l_{21}\} , \{\mybar{ l_{11}}, \mybar{ l_{21}}\}\}$,  the following derivation is produced: 
\vspace*{3mm}
\begin{center}
\AxiomC{$ $}
\RightLabel{$\Conflict$}
\UnaryInfC{$l_{11}, l_{21} \vdash \emptyset$}
\RightLabel{$\Red$}
\UnaryInfC{$ l_{11}, l_{21} \vdash  \{ \mybar{ l_{21}} \}$}
\RightLabel{$\Red$}
\UnaryInfC{$ l_{11}, l_{21} \vdash  \{ \mybar{ l_{11}} , \mybar {l_{21}} \}$ }
\RightLabel{$\Unit$}
\UnaryInfC{$ l_{11} \vdash  \{l_{21} \}, \{ \mybar{ l_{11}} , \mybar{l_{21}} \} $}
\RightLabel{$\Unit$}
\UnaryInfC{$ \vdash \{ l_{11} \}, \{l_{21} \}, \{ \mybar{ l_{11}} , \mybar{ l_{21}} \} $}
\DisplayProof
\end{center}
\vspace*{3mm}
The following is the Minlog output for the pigeon hole formulae $\mathbf{PHP}(2,1)$. There is a constructor \verb|CsuccessZero| of the algebra \verb|success| which represents the disjunction in the main proof statement. The data type extracted from this algebra can be seen as a union type that contains either a DPLL derivation or a model of the formula. In this case it contains a DPLL derivation showing that the formula is unsatisfiable. The arguments to \verb|CsuccessZero| store how the $\Conflict$ is derived. The literal $l_{11}$ is represented as \verb|(Pos(Variable 11))| in the Minlog formalization and the clause $\{ \mybar{ l_{11}} , \mybar{l_{21}} \}$ is represented as \verb|CC(Neg(Variable 21)::(Neg(Variable 11)):)|.

\begin{verbatim}
        CsuccessZero
         (CUnit
          (Pos(Variable 11))
          (CUnit
           (Pos(Variable 21))
           (CRed
            (CC(Neg(Variable 21)::(Neg(Variable 11)):))
            (Pos(Variable 21))
            (CRed
             (CC(Neg(Variable 11)):)
             (Pos(Variable 11))
             CConflict))))
\end{verbatim}

Running the DPLL solver on a satisfiable formula results in a function which 
maps literals to booleans. For example running the solver with 
$\mathbf{PHP}(2,2)$ results in the function  
	 $\mathrm{M}: \mathrm{literals} \to \mathbb{B}$  where  
$\mathrm{M}(l) = \True$ iff  $l \in \{ l_{12}, \mybar{ l_{11}}, l_{21}, \mybar{ l_{22}}  \}$.
The Minlog output for the satisfiable formula $\mathbf{PHP}(2,2)$ is as follows. Here the square brackets represent a lambda abstraction for the literal $l_0$. The model $M$ is written as $\lambda l_0 . l_0 \in \{ l_{12}, \mybar{l_{11}},l_{21}, \mybar{l_{22}} \}$.
\begin{verbatim}
        CsuccessOne
         ([l0]
          [if (l0=Pos(Variable 12))
              True 
              [if (l0=Neg(Variable 11)) 
                  True 
                  [if (l0=Pos(Variable 21)) 
                      True 
                      (l0=Neg(Variable 22))]]])
\end{verbatim}

\subsection{Comparison of Program Performance}
\label{sec:compare-performance}
\begin{table}
\caption{Performance in Minlog versus Haskell}
\label{tab:minlog-vs-haskell}
\begin{center}
{\small
\begin{tabular}{ccccccc}
  \toprule
  \text{Formula} & \text{Minlog $\forall$}  & \text{Minlog $\forallnc$} & \multicolumn{2}{c}{\text{Compiled (\texttt{ghc -O2})}} & \multicolumn{2}{c}{\text{Compiled (\texttt{ghc -O2 -fllvm})}} \\ \cmidrule(r){4-5} \cmidrule(l){6-7}
  & \text{Witness} & \text{Witness} &  \text{Witness} & \text{Yes/No} & \text{Witness} & \text{Yes/No} \\ \midrule
  PHP(4,3) & 33.62s & 11.61s    & 0.019s   & 0.006s  & 0.015s & 0.004s  \\
  PHP(4,4) &  5.45s  &   5.25s      &  0.019s   & 0.010s & 0.014s & 0.007s   \\ 
  PHP(5,4) &  13m54s      &  2m41s   & 0.055s   &  0.020s & 0.036s & 0.012s  \\
  PHP(5,5) & 26.09s &  25.03s    &  0.024s   &  0.015s & 0.020s & 0.010s \\
  PHP(6,5) & 5h35m41s & 37m25s  &  0.367s  & 0.066s & 0.279s & 0.039s \\
  PHP(6,6) &    1m34.11s    &  1m24.88s    &  0.035s  & 0.025 & 0.025s & 0.015s    \\ \addlinespace
  PHP(8,8) &   - & - & 0.054s & 0.029s & 0.040s & 0.025s \\
  PHP(9,8) &   - & - & -  & 1m21.915s & - & 32.062s  \\ 
  PHP(9,9) &   - & - &  0.064s & 0.042s & 0.052s & 0.030s \\
  PHP(10,9) &   - & - & - & 102m 16s  & - & 15m 5s \\ \bottomrule
\end{tabular}
}
\end{center}
\end{table}

%
%
The $\forall$ solver and $\forallnc$ solver were compared using both unsatisfiable  $\mathbf{PHP}(n+1, n)$ and satisfiable $\mathbf{PHP}(n,n)$ pigeon hole formulae. The unsatisfiable pigeon hole formulae are harder than the satisfiable formulae as they have a large search space that must be traversed entirely by the solver in order to construct a derivation.
This difficulty can be seen -- compare column 2 and 3 in Table 
\ref{tab:minlog-vs-haskell} 
-- when both the $\forall$ and $\forallnc$ solver are applied to the unsatisfiable pigeon hole formulae. The solver without the optimization takes considerably longer to construct a derivation of unsatisfiability. This is due to computationally irrelevant data being stored in the unoptimized derivations. \medskip \\ 
%
%
The next two columns of Table \ref{tab:minlog-vs-haskell} present two versions of the $\forallnc$ solver when extracted to Haskell and compiled by the Glasgow Haskell Compiler (GHC). The first 
returns a witness of the result i.e.\ either a model which satisfies the formula or a derivation of its unsatisfiability. 
The second 
returns only a Yes or No answer as to whether a formula is satisfiable or not. Due to the inherent laziness of Haskell 
the two programs differ quite dramatically in their behavior. The solver that returns a Yes/No answer performs considerably faster compared to the solver which produces the witness in addition. By using the Low Level Virtual Machine (LLVM) backend~\cite{CL04} for GHC
, a further speed up was achieved, which can be seen in the last two columns of Table \ref{tab:minlog-vs-haskell}.

%

%
%
%


\begin{table}
\caption{Performance compared to Versat}
\label{tab:versat}
\begin{center}
{\small
\begin{tabular}{ccc}
  \toprule
  \text{Formula} & \text{$\forallnc$ compiled (Yes/No)}  & \text{Versat} \\ \midrule
  PHP(7,6) &   0.226s & 0.089s \\
  PHP(8,7) &   2.42s & 0.794s \\ 
  PHP(9,8) &   32.062s & 17.217s \\ 
  PHP(10,9) &  15m 5s & 15m 46s \\ \hline
\end{tabular}
}
\end{center}
\end{table}

We also compared the performance of our $\forallnc$ solver, compiled using the LLVM backend of GHC, with that of Versat~\cite{DO12}. Our solver was run with the option of not computing a witness since Versat does generally not compute a proof. The results in Table \ref{tab:versat} show that our solver is comparable with Versat. It is slower on the easier formulae and faster on the hardest pigeon hole formulae. This is because the clause learning optimization of Versat has some overhead and does not increase the performance on pigeon hole formulae. The point of the learned clauses is to reduce the search space for the solver.  In this case, they instead consume more memory and time to compute.

\begin{table}
\caption{Industrial case study: Extracted solver versus Versat}
\label{tab:industrial}
\begin{center}
{\small
\begin{tabular}{ccc}
  \toprule
  \text{Formula} & \text{$\forallnc$ compiled (Yes/No)}  & \text{Versat} \\ \midrule
  $\neg R1 \vee \neg R3$ & 7.028s & 0.050s \\
  $\neg R1 \vee \neg R4$ & 6.961s & 0.040s \\
  $\neg R2 \vee \neg R3$ & 7.105s & 0.053s \\
  $\neg R2 \vee \neg R4$ & 7.059s & 0.044s \\
  $\neg R3 \vee \neg R4$ & 7.015s & 0.047s \\
\end{tabular}
}
\end{center}
\end{table}

\subsection{Industrial Case Study}
The same version of our solver was also applied to the verification of a real world railway control system which was provided by our industrial partner Invensys Rail (now Siemens), via a description in Ladder logic. We adapted \cite{KK08} to translate  Ladder logic programs into Minlog/Haskell and the industrial tool  SCADE~\cite{SCADE}, and also performed a comparison with Versat.
The SAT problem is formulated to perform falsification checking, as described in \cite{MS00}, that is, a satisfying assignment represents a counter example, and an unsatisfiable result means the safety property can not be violated in the system. 
The size of our case study is 14726 clauses and 8166 variables. 
For comparison, we present the run-times for checking five safety conditions which show that two conflicting routes, out of a set of four routes $R1, \ldots, R4$, can not be active in the railway at the same time.  For each of the five conditions our solver produces a proof certifying that the safety property holds in approximately 7s.  The SCADE suite can verify that each of the safety properties holds in less than one second 
(no greater accuracy of run-times provided by the system for this case). 


While we cannot expect to compete with an industrial tool on speed and
functionality, we have been able to solve a large practical problem in
a reasonable amount of time. It is important to note that the solver
inside the SCADE suite has not been formally verified whereas our
solver has. Interestingly however, also Versat solves these problems in less
than one second -- see Table~\ref{tab:industrial} for a comparison between our 
extracted solver and Versat -- that is, we may conclude
that optimizations such as clause-learning and the use of efficient data
structures that enable to efficiently parse and identify (un-)satisfiability 
of a formula indeed improve the 
performance for this type of problems (and our extracted solver
should be extended by these optimizations as well).

\section{Conclusion}

We have presented a conceptually new approach to the synthesis and
verification of SAT algorithms that, in contrast to similar work in Coq and Isabelle \cite{SL08,FM10} does not require the formalization
of the SAT programs in the formal system, but obtains SAT algorithms purely by program extraction. To this end, we formalized the DPLL proof system and
performed a constructive proof from which a correct SAT solving
algorithm was extracted automatically. The extracted program attempts
to show the (un)satisfiability of a propositional formula in
conjunctive normal form. If the CNF formula is satisfiable it produces
a model of the formula; otherwise it produces a derivation showing
unsatisfiability. We strategically placed $\forallnc$ quantifiers into
the proof to reduce the complexity of the extracted program and
increase its performance.  The solver containing $\forallnc$
quantifiers was extracted into the functional programming language
Haskell, and the performance of the two solvers was evaluated using
pigeon hole formulae.
We have also shown how it is possible to extract a
program that translates between DPLL and resolution proofs. This was
done in such a way that we obtain some qualitative information about
quantitative aspects of the extracted program i.e. computational
complexity. Using this translation it was possible to extract a
resolution solver based on the DPLL proof system.

Overall, our paper shows that the approach of developing verified
programs via extraction from proofs is scalable to non-trivial
applications.  Furthermore, it demonstrates how to include efficiency
considerations into this approach. For instance, we have avoided
repeated unnecessary look-ups of clauses by the split of clause sets
in two sets $\Delta$ and $\Theta$. This counters the often heard
argument that with program extraction one 'loses the grip' on the
program and its efficiency. It is important to note that these
efficiency considerations do not compromise the correctness of the
extracted program since these are applied at the proof level where
correctness is guaranteed by the proof system.

We consider the fact that our approach does not require any formalization 
of algorithms a major advantage, since it means that
program development via extraction can be carried out in a formal system
that is much more lightweight than in the verification approach, where
the term language must include a programming language, and the meaning 
of the programming constructs must be specified by axioms and proof rules. 
This advantage is particularly striking in applications in 
analysis~\cite{UB11,UB12} where corecursive exact real number 
algorithms (whose formalization and
specification is non-trivial and subject of ongoing research) can be 
automatically extracted from proofs involving only coinductive definitions
in the form of largest fixed points of predicate transformers.

\subsection{Future Work}
There are two directions for further work: applying our method to extract a more advanced class of SAT solvers, and applying our approach to a different class of decision problems.

We are in the process of formalizing optimizations such as clause learning and conflict analysis \cite{Biere2009,MM01,MS99}.  This requires a modification of the DPLL proof system such that it captures the additional behavior. A completeness theorem has then been proven for the modified calculus. We currently have extracted a prototype clause learning solver from this proof. In order for this solver to be an improvement on the previous one we need to lower the computational overhead resulting from  clause learning. Such a solver would also benefit from lazy data structures such as the two-watched-literal scheme. It is unclear whether the inherent laziness of Haskell will provide the same effect as these data structures or if they would have to be formalized as part of the proof.

It is desirable to be able to solve not just propositional formulae but also first-order formulae. This is possible by extending SAT algorithms so that they can apply some background theory for first order formulae. Such algorithms are called Satisfiability Modulo Theories (SMT) solvers. We would have to formalize a proof system used by SMT solvers such as abstract DPLL \cite{RN05} and then perform a completeness proof. A solver extracted from such a proof system would be able to solve a broader range of problems described in a language richer than propositional logic.

\subsection{Sources}
The Minlog formalization optimized with $\forall_{nc}$ quantifiers and its extracted program as Haskell code can be found at \url{http://cs.swan.ac.uk/minlog/dpll/}.

\subsection*{Acknowledgment} 
We would like to thank the anonymous referees for their constructive comments.
The financial support of Invensys Rail/Siemens Rail Automation is gratefully
acknowledged.


\bibliographystyle{plain}
\bibliography{lit}

\end{document}